\documentclass[a4paper]{article}
\usepackage{epsfig}

\begin{document}

\twocolumn[

{\small\it
Solar System Research,
Vol.~32, No.~4, 1998, pp.~323--325.\\
Translated from Astronomicheskii Vestnik,
Vol.~32, No.~4, 1998, pp.~367--369.\\
Original Russian Text Copyright \copyright 1998 by Dumin\\

}

\hrule
\vspace{0.5ex}
\hrule

\vspace{6ex}

\begin{center}
{\Large\bf On the Physical Nature of the Magnetic-Field\\
Freezing-in Effect in Collisionless Cosmic Plasmas

}

\bigskip
\smallskip

{\large\bf Yu.~V.~Dumin}

\medskip

{\small
{\it Institute of Terrestrial Magnetism, Ionosphere,\\
and Radiowave Propagation,\\
Russian Academy of Sciences,\\
Troitsk, Moscow oblast, 142092 Russia}

\bigskip

Received January 15, 1998

}

\end{center}

\smallskip

\begin{quotation}

{\small
\noindent
{\bf Abstract}---Attention is drawn to a fundamental difference
between the physical mechanisms responsible for the magnetic field
freezing-in effect in spatially isotropic (collisional) and
strongly anisotropic (collisionless) plasmas. (The first case is
characteristic of internal regions of the Sun and other stars;
the second, for the solar corona, interplanetary and interstellar
medium, planetary ionospheres and magnetospheres, cometary tails.)
It is shown that particles of a collisionless plasma remain in the
same field line during their motion because of the properties of
electrodynamic drift which stem from the divergence-free character of
the magnetic field and the equipotentiality of the field lines
(but not as a result of the appearance of induction currents,
compensating variations of the external magnetic flux through
a closed contour, as in the case of isotropic conductivity of
the plasma).

}

\end{quotation}

\bigskip

]

\begin{center}
INTRODUCTION
\end{center}

\smallskip

The magnetic-field freezing-in theorem is well known and widely used
in various branches of plasma physics and, particularly, for the
description of plasma phenomena in space over several decades. This theorem
is applied both to plasmas with high conductivities in each of the three
directions (which corresponds to the case of a collision frequency large
in comparison with the gyrofrequencies of charged particles) and to plasmas
whose conductivity is high only in the magnetic-field direction, being
close to zero in the two other directions (which corresponds to small
collision frequencies). The first case takes place in the internal
regions of the Sun and other stars, whereas the second case is
characteristic of the solar corona, interplanetary and interstellar medium,
planetary ionospheres and magnetospheres, and cometary tails.

Although the collisionless case is, in a certain sense, more typical of
astrophysical conditions, in current publications that present
a mathematical treatment or, all the more so, a descriptive interpretation
of the magnetic-field freezing-in effect, specific features of this
phenomenon in collisionless plasmas are either overlooked completely or
treated incorrectly by many authors.

As is known (Pikel'ner, 1961), the mathematical statement of
the freezing-in theorem is given by the condition

$$
d \, \Phi / d \, t \, = \, 0 \: ,
\eqno {\rm (1)}
$$
where $ d / dt $ denotes the total derivative with respect to time,
and $ \Phi $ is the flux of the magnetic field $ {\bf B} $ through
a closed contour bound to the moving medium:
$$
\Phi \, = \int\!\!\!\int {\bf B} \: d {\bf S} \; .
\eqno {\rm (2)}
$$
(Hereinafter, the Gaussian system of units is used throughout.)

An illustrative description of this effect in the case of isotropic
conductivity (Sivukhin, 1977) is based on the fact that the electric
induction current $ I_{\rm ind} $ generated by the motion of the closed
contour is determined by the expression
$$
I_{\rm ind} =  \, - \, \frac{1}{cR} \, \frac{d \, \Phi}{d \, t} \; ,
\eqno {\rm (3)}
$$
where $c$ is the speed of light in vacuum, and $R$ is the total resistance
of the contour. If $R$ tends to zero, then the condition of boundedness
of the induction current $ I_{\rm ind} $
implies the conservation of the magnetic flux $\Phi$
in the course of time, which is expressed by the freezing-in
condition (1). In other words, the induction current in the contour,
arising according to the Lenz rule, is so large that it can completely
compensate the magnetic-flux variation caused by external sources.
Consequently, plasma particles are free to travel along the magnetic field
but must always remain in the same field line.

\begin{figure}[thb]
\centering
\epsfig{file=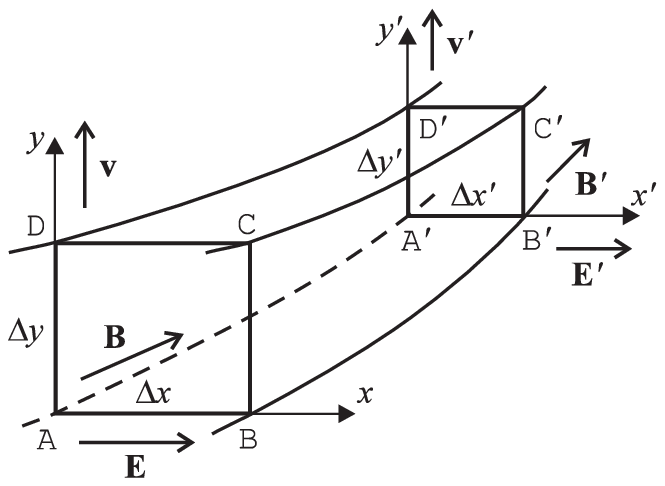}
\end{figure}

On the other hand, it is clear that the above consideration of a perfectly
conducting closed contour is possible only for a plasma with a high
conductivity in any direction. In the case of a strongly anisotropic
plasma (with a high conductivity only along the magnetic field),
this procedure is meaningless.

The need for special consideration of a strongly anisotropic plasma was
emphasized over 40 years ago in the well-known review by
Syrovatskii (1957), devoted to the magnetohydrodynamics of media with
isotropic conductivity. Nevertheless, up to now, most authors either
restrict the proof of the freezing-in theorem to only the case of isotropic
conductivity or derive the condition (1) for a strongly anisotropic plasma
by means of formal transformations of the equations of its motion.
The last approach is quite correct from the mathematical point of view,
but it cannot clarify the physical mechanism of the freezing-in effect in
the collisionless case. It is this problem that is the central topic
of the present paper.

\bigskip
\smallskip

\begin{center}
THE FREEZING-IN MECHANISM\\
IN A COLLISIONLESS PLASMA
\end{center}

\smallskip

First of all, it should be emphasized that when the plasma is cold and
the frequencies of collisions between its particles are considerably
less than the gyrofrequencies, then the macroscopic motion of the plasma
can be caused only by the ${\bf E}\!\!\times\!\!{\bf B}$-drift.
The reason is that pressure forces are negligibly small in such conditions,
while all other kinds of charged-particle drift in the magnetic field
are possible for only a nonzero thermal velocity. For example,
the centrifugal drift requires a sufficiently large particle velocity
along the magnetic field, whereas the drift caused by the magnetic-field
nonuniformity in a transverse direction requires a nonzero gyroradius and,
consequently, the respective nonzero component of the thermal velocity.
(It should be mentioned that this relation between the
${\bf E}\!\!\times\!\!{\bf B}$-drift and the magnetic field freezing-in
effect in a collisionless plasma has been pointed out several times
in previous works, starting from the pioneering studies by
Alfv\'en and F\"althammar (1963) up to the present time
(Filippov, 1997); however, such treatments were usually carried out
only on a formal mathematical basis.)

To represent pictorially the physical mechanism of the phenomenon
under consideration, let us consider a fairly thin flux tube
of the magnetic field ${\bf B}$ with a rectangular cross section
(see figure). Let $ {\tt ABCD} $ and $ {\tt A'B'C'D'} $ be two arbitrary
cross sections of this tube. All physical quantities characterizing
the plasma will from now on be written with or without primes, depending
on the cross section they refer to.

Let the particle under consideration be located at point $ {\tt A} $
at the initial instant of time. As already pointed out,
its macroscopic motion across the magnetic field can be described,
to a first approximation, as a drift with a velocity ${\bf v}$
under the action of the electric field ${\bf E}$, which is assumed to be
directed along the $x$-axis. Consequently, in the time $ \Delta t $
this particle will be displaced along the $y$-axis by the distance
$$
\Delta y \, = \, c \, (E/B) \, \Delta t
\eqno {\rm (4a)}
$$
(we do not need the vector product in the calculation of the
${\bf E}\!\!\times\!\!{\bf B}$-drift,
since in the geometry at hand the vectors ${\bf v}$, ${\bf E}$,
and ${\bf B}$ are mutually orthogonal).

Similarly, another particle, initially located at the point
$ {\tt A'} $ in the same field line as the point $ {\tt A} $,
will be displaced by the distance
$$
\Delta y \hspace{0.1em} ' =
\, c \, (E \hspace{0.05em} ' / B \hspace{0.05em} ') \, \Delta t \; .
\eqno {\rm (4b)}
$$

Further, if the electrical conductivity is assumed to be infinitely large
along the magnetic field lines and zero across them, then the field lines
should be equipotential. Consequently, the electric-field
strength at sections $ {\tt ABCD} $ and $ {\tt A'B'C'D'} $
can be represented in the form:
$$
E \, =
\, - \, \Delta \varphi / \Delta x \; ,
\eqno {\rm (5a)}
$$
$$
E \hspace{0.05em} ' \, =
\, - \, \Delta \varphi / \Delta x \hspace{0.05em} ' \; ,
\eqno {\rm (5b)}
$$
where
$ \Delta \varphi \, = \, \varphi ({\tt B}) - \varphi ({\tt A}) \, =
\, \varphi ({\tt B'}) - \varphi ({\tt A'}) $
is the potential difference between the field lines under consideration.

It is very important to note that, although points $ {\tt B} $ and
$ {\tt B'} $ were by definition taken to be located in the same field
line, in general, this does not imply that the displacements
$ \Delta y $ and $ \Delta y \hspace{0.1em} ' $ of the particles by the
${\bf E}\!\!\times\!\!{\bf B}$-drift will lead to new positions
$ {\tt D} $ and $ {\tt D'} $ located again in the same field line.

To establish whether or not this actually takes place, let us substitute
the expressions (5a) and (5b) for the electric fields into formulas (4a)
and (4b) respectively and then divide one of them by the other.
As a result, we find that the displacements $ \Delta y $ and
$ \Delta y \hspace{0.1em} ' $ are related as follows:
$$
B \, (\Delta x) \, (\Delta y) \, =
\, B \hspace{0.05em} ' \, (\Delta x \hspace{0.05em} ') \,
(\Delta y \hspace{0.1em} ') \; .
\eqno {\rm (6)}
$$

On the other hand, expression (6) is the condition of the conservation
of the magnetic flux through cross sections $ {\tt ABCD} $ and
$ {\tt A'B'C'D'} $. Because of the divergence-free character of the
magnetic field ($ {\rm div} {\bf B} = 0 $), this condition is true
if and only if point $ {\tt D'} $ belongs to the same field line as
point $ {\tt D} $ (i.e., surface $ {\tt DD'C'C} $ specifies the upper
boundary of the flux tube). Therefore, all particles located at the
initial instant of time in the field line $ {\tt AA'} $ will pass
to one field line $ {\tt DD'} $ in the time interval $ \Delta t $,
which is the required result.

\bigskip
\smallskip

\begin{center}
CONCLUSION
\end{center}

\smallskip

Therefore, as shown in the present paper, although the magnetic field
freezing-in effect is universal (i.e., manifests itself in a similar way
in media with isotropic and strongly anisotropic conductivities),
the physical mechanisms responsible for it are absolutely different
in these two limiting cases. In an isotropic plasma, freezing-in is due
to induction currents, which compensate the variation of the magnetic
flux through a closed contour moving together with the conducting
substance. In the case of a plasma with a strong anisotropy, freezing-in
is a direct consequence of particularities of the electrodynamic drift
which are associated with the divergence-free character of the magnetic
field and the equipotential character of the field lines.

\bigskip
\smallskip

\vspace{15cm}

\begin{center}
ACKNOWLEDGEMENTS
\end{center}

\smallskip

I would like to thank
V.I.~Badin,
M.G.~Deminov,
M.A.~Livshits,
B.V.~Somov, and
L.M.~Svirskaya
for helpful comments
and to all participants of the IV~Congress of the Euro--Asian Astronomical
Society who took part in the discussion of my report.

\bigskip
\smallskip

\begin{center}
REFERENCES
\end{center}

{\small
\tolerance=2000

\begin{description}

\item[Alfv\'en,~H. and F\"althammar,~C.-G.,]
{\it Cosmical Electrodynamics: Fundamental Principles,}
Oxford: Clarendon, 1963, p.~191.
Translated under the title
{\it Kosmicheskaya elektrodinamika: Osnovnye prin\-tsipy,}
Moscow: Mir, 1967.

\item[Filippov,~B.P.,]
Two-Ribbon ${\rm H}_{\alpha}$-Emission during the Eruption of a Filament,
{\it Astron.\ Zh.,}
1997, vol.~74, no.~4, pp.~635--640.

\item[Pikel'ner,~S.B.,]
{\it Osnovy kosmicheskoi elektrodina\-miki}
(Fundamentals of Cosmic Electrodynamics),
Mos\-cow: Gos.\ Izd.\ Fiz.--Mat.\ Lit., 1961, pp.~58--69.

\item[Sivukhin,~D.V.,]
{\it Obshchii kurs fiziki, Tom III: Elektrichestvo}
(General Course of Physics, vol. III: Electricity),
Moscow: Nauka, 1977, pp.~295--296.

\item[Syrovatskii,~S.I.,]
Magnetohydrodynamics,
{\it Usp.\ Fiz.\ Nauk,}
1957, vol.~62, no.~3, pp.~247--303.

\end{description}

}

\end{document}